\documentclass[aps,prb, groupedaddress, superscriptaddress] {revtex4}
\usepackage{epsfig}
\usepackage{booktabs}

\usepackage{graphicx}
\usepackage{dcolumn}
\usepackage{bm}
\usepackage{mathrsfs}
\begin{document}


\title{How electrons Coulomb repulsion changes Graphene band structure}
\author{Rostam Moradian}
 \email{rmoradian@raz.ac.ir}
\author{ Poorya Rabibeigi}
\affiliation{ Physics Department, Faculty of Science Razi University, Kermanshah, Iran}.
\affiliation{Nano science and nano technology research center,  Razi University, Kermanshah, Iran.}


\date{\today}

\begin{abstract}
How electron-electron Coulomb repulsion modifies electronic band structure is a big change in strongly correlated systems. We introduced a method for calculation of realistic band structure of these systems and eliminating fake states. By using this method we investigated how electrons repulsion renormalizes graphene band structure. Our results show that in the dynamical mean field theory at Coulomb repulsion $u=2.596 t$ a four band semi metal to four bands anti ferro magnetism phase transition starts while due to inter site correlation for four sites, $nc=4$, effective medium super cell approximation it is $u=3.2 t$.

\end{abstract}

\pacs{Valid PACS appear here}
\maketitle

Strongly Correlated materials cover variety of notable properties such as unconventional superconductivity to metal-insulator transitions. Although theoretical calculations of these systems properties are hard but their applications is wide\cite{Kotov}. For a honeycomb lattice by using Hubbard model with on site electrons repulsion and a finite size quantum Monte Carlo calculations Sorella at $u/t>4.5$ a non magnetic semimetalic to an anti ferromagnetic insulator phase transition found\cite{Sorella1992, Martelo, Paiva}. Also by same model and method Meng et al.\cite{Meng} reexamined the ground state phase diagram of graphene like systems. They suggested a spin liquid phase between semi metal at $u/t<3.4$ and anti ferromagnet phase at $u/t>4.3$ that is in the range $3.4<u/t < 4.3$. By extended to higher cluster sizes Sorella at. al\cite{Sorella2012} showed no spin liquid phase for these systems and a direct semimetalic to an anti ferromagnetic insulator phase transition happens.  Effect of electrons repulsion on bilayer graphene dispersion within the Hartree-Fock-Thomas-Fermi theory studied\cite{Kusminskiy}. Tang et. al by using random phase approximation showed that electron-electron interactions in two-dimensional Dirac fermion systems leads to two previously predicted regimes: a Gross-Neveu transition to a anti ferromagnetic Mott insulator and a semi metallic state\cite{Tang}. Recently by proximity screening controlling of electron-electron repulsion in graphene investigated\cite{Kim}. All graphene like physical quantities that mentioned could be calculate if one have renormalized band structure. In the average system the electron-electron interaction effect is inside of self energy $\Sigma({\bf k}, E)$ that  changes non interacting band structure $E^{0}_{\bf k}$ to renormalized band structure $E_{\bf k}$. Self energy could not be calculate exactly.  One of main problem in condensed matter physics is choosing  type of self energy approximation. For weak interaction Hatree Fock approximation that is a mean field approximation widely used\cite{Stauber}. Dynamical mean field theory (dmft) is a single site approximation that could be use for any interaction strength $u$\cite{Vollhardt, Georges}. In the dmft inter sites correlation is neglected.  Multi sites, $Nc$, dynamical mean field approximation (dca) for including multi sites correlation is introduced\cite{Hettler1998,Hettler2000,Jarrell2001, Jarrell2001-2}. They claimed that dca recovers exact self energy $\Sigma({\bf k}, E)$ in the limit of $limit_{nc\rightarrow\infty} \mathcal{N}$. The dca relation between k-space and real space self energies defined by $\Sigma(I,J;E)=\sum_{{\bf K}_{n}}\Sigma({\bf K}_{n}, E) e^{i{\bf K}_{n}.{\bf r}_{IJ}}$ and $\Sigma({\bf K}_{n};E)=\frac{1}{Nc}\sum_{IJ}\Sigma(I, J, E) e^{-i{\bf K}_{n}.{\bf r}_{IJ}}$\cite{Hettler1998,Hettler2000,Jarrell2001, Jarrell2001-2} where in limit $Nc\rightarrow\ \mathcal{N}$ that does not recover exact k-space and real space self energies $\Sigma(i,j; E)=\frac{1}{\mathcal{N}}\Sigma({\bf k}; E)e^{i{\bf k}.{\bf r}_{ij}}$ and   $\Sigma({\bf k}, E)=\frac{1}{\mathcal{N}}\Sigma(i,j; E)e^{-i{\bf k}.{\bf r}_{ij}}$. To fix these problems we introduced effective medium super cell approximation (emsca)\cite{Moradian01,Moradian02,Moradian03}. We showed that neglecting k-space whole contributions of super cells self energies in different super cells leads to step function coarse grained self energies in the first Brillouin zone and periodicity of real space self energy with respect to super cell lengths\cite{Moradian03}. In the emsca $\Sigma(I,J;E)=\frac{1}{Nc}\sum_{{\bf K}_{n}}\Sigma({\bf K}_{n}, E) e^{i{\bf K}_{n}.{\bf r}_{IJ}}$ and $\Sigma({\bf K}_{n};E)=\frac{1}{Nc}\sum_{IJ}\Sigma(I, J, E) e^{-i{\bf K}_{n}.{\bf r}_{IJ}}$ where in limit $Nc\rightarrow\ \mathcal{N}$ both of them recovers exact k-space and real space self energies. To eliminate discontinuities of k-space self energies $\Sigma({\bf K}_{n}, E)$ in the FBZ we introduces another relation $\Sigma({\bf k};E)=\frac{1}{Nc}\sum_{IJ}\Sigma(I, J, E) e^{-i{\bf k}.{\bf r}_{IJ}}=\frac{1}{Nc^{2}}\sum_{IJ}\sum_{{\bf K}_{n}}\Sigma(I, J, E) e^{i({\bf K}_{n}-{\bf k}).  {\bf r}_{IJ}}$ which recovers both deft and exact self energies. Another main problem of approximated self energies is creating fake electronic states that should be eliminate. Now we apply our method by eliminating fake states to obtain  correlated graphene realistic band structure that is a big challenge in condensed mater physics.    

We start our investigation by a Hubbard model for a strongly correlated system which is given by,       
\begin{eqnarray}
H&=&-\sum_{ij\sigma\sigma}t^{\alpha\beta}_{ij}{c^{\alpha}_{i\sigma}}^{\dagger}c^{\beta}_{j\sigma}+\sum_{i\alpha}u \hat{n}^{\alpha}_{i\uparrow}\hat{n}^{\alpha}_{i\downarrow}-\sum_{i\sigma} \mu { c^{\alpha}_{i\sigma}}^{\dagger}c^{\alpha}_{i\sigma}
\label{eq:Hamiltonian}
\end{eqnarray}
where ${ c^{\alpha}_{i\sigma}}^{\dagger}$ ($c^{\alpha}_{i\sigma}$) is the creation (annihilation) operator of an electron with spin $\sigma$ on $\alpha$ subsite of lattice site $i$ and $\hat{n}^{\alpha}_{i\sigma}={c^{\alpha}_{i\sigma}}^{\dagger}c^{\alpha}_{i\sigma}$ is the electrons number operator. $t^{\alpha\sigma\beta\sigma}_{ij}$ are the hopping integrals between $\alpha$ subsite of $i$ and $\beta$ subsite of $j$ lattice sites. $\mu$ is the chemical potential.

\begin{figure}
\centerline{\epsfig{file= 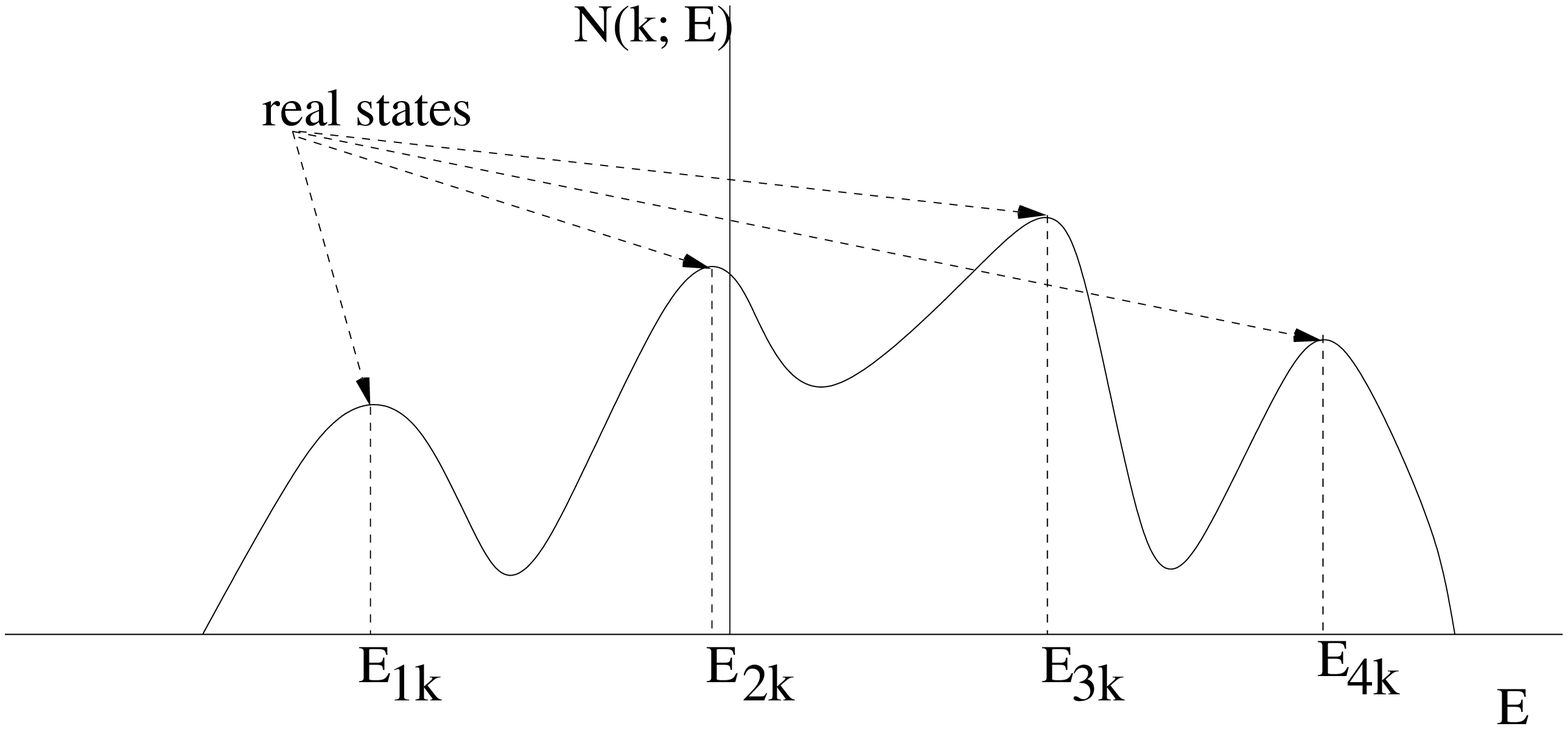 ,width=6.0cm,angle=0}}
\caption{shows real states eigen values of calculated density of states are maximums of $N({\bf k}; mE)$. Other states are fake that should be eliminate.   }
 \label{figure:real-dos} 
\end{figure}

\begin{figure}
\centerline{\epsfig{file= 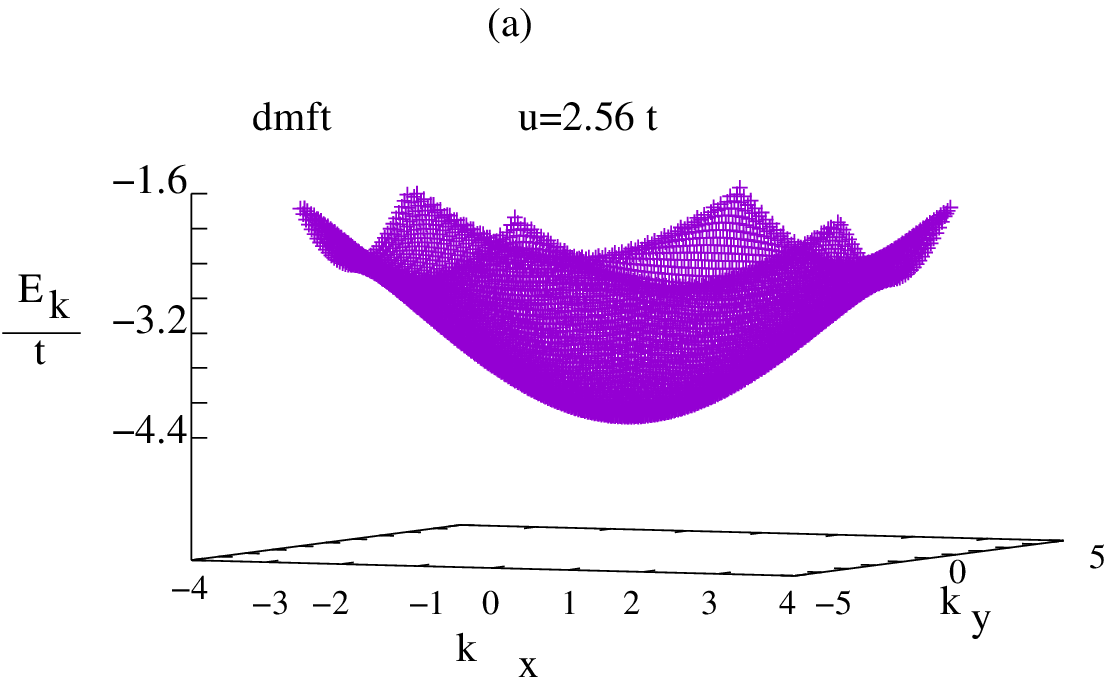 ,width=6.0cm,angle=0}\epsfig{file=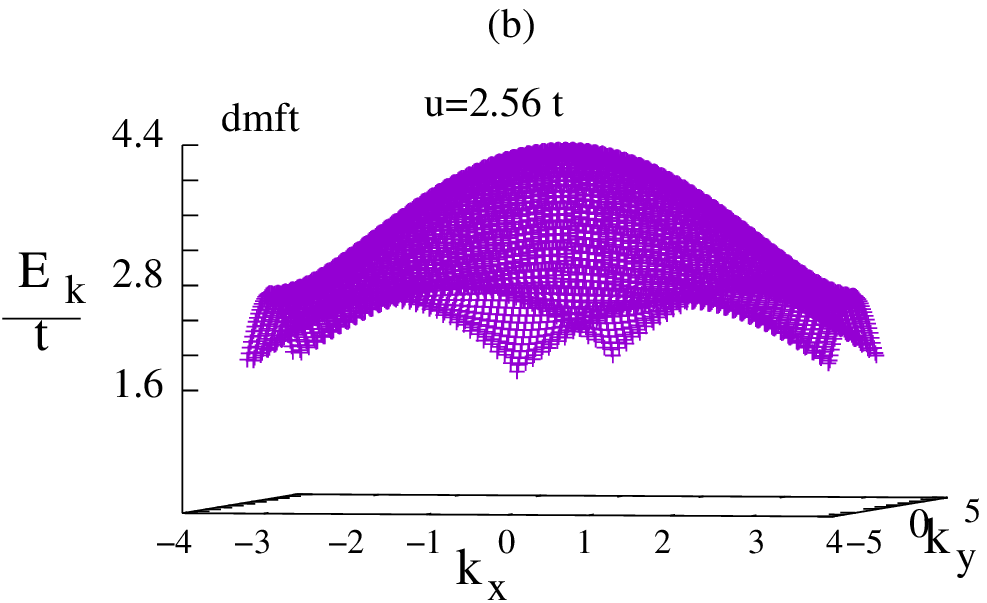  ,width=6.0cm,angle=0}}
\centerline{\epsfig{file= 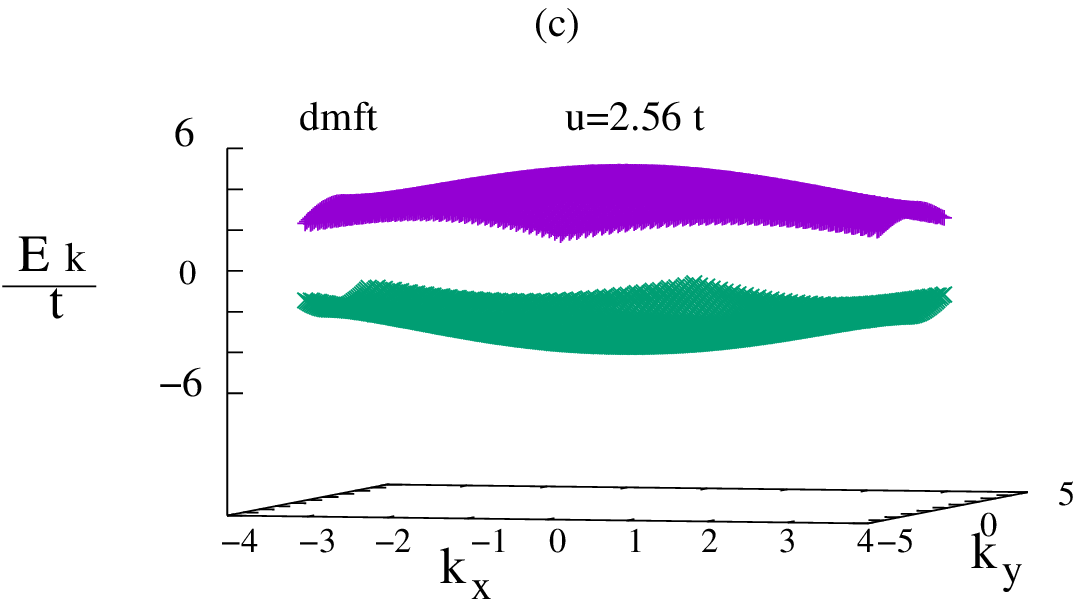,width=6.0cm,angle=0}\epsfig{file=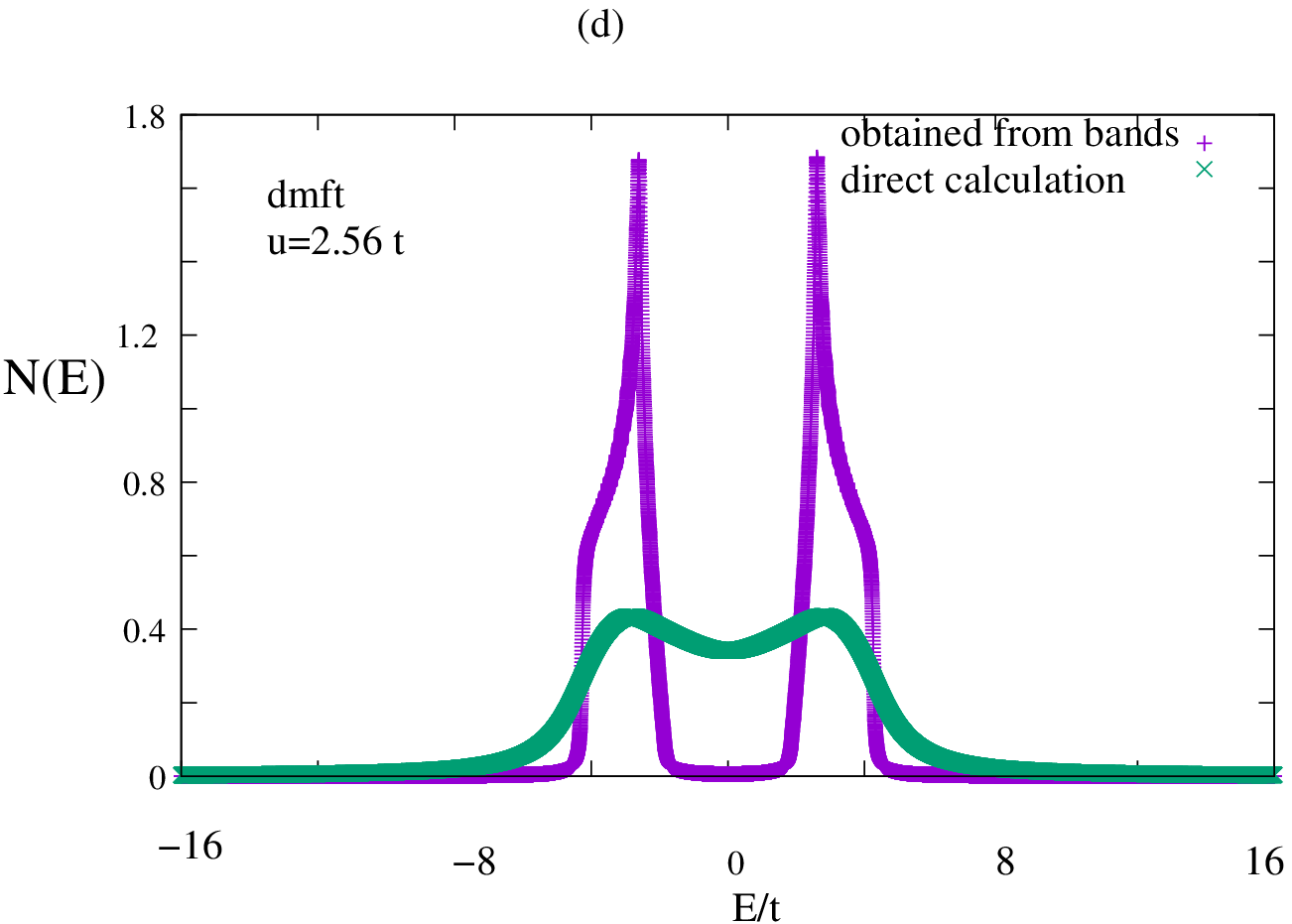  ,width=6.0cm,angle=0}}
\caption{(Color on line) (a), (b) and (c) shows graphene real valance, conduction and both valance and conduction bands for $u=2.56 t$ respectively. In this repulsion potential valance and conduction bands are separated but spin up and down bands still are not separated. (d) shows comparison of dos obtained from real calculated bands with dos obtained from dmft local Green function. There many fake states in the dmft dos. }
 \label{figure:dmft-u2-56} 
\end{figure}

The equation of motion for electrons corresponding to the above Hamiltonian, Eq.\ref{eq:Hamiltonian}, is given by, 
\begin{eqnarray}
\sum_{l\gamma }\left((-{\bf I}\frac{\partial}{\partial\tau}+\mu{\bf I})\delta_{il}\delta_{\gamma\alpha}+{\bf t}^{\alpha\gamma}_{il}\right){\bf G}^{\gamma\beta}(l,j;\tau)-\sum_{\gamma}{\bf G}_{2}(i\alpha,i\gamma,j\beta;\tau)
=\delta(\tau)\delta_{ij}\delta_{\alpha\beta}{\bf I}
\label{eq:normal-general equation of motion}
\end{eqnarray}
where $\bf{I}$ is a $2\times 2$ unitary matrix, ${\bf t}^{\alpha\beta}_{ij}=t^{\alpha\uparrow\beta\uparrow}_{ij}{\bf I}$ and two particle Green function ${\bf G}_{2}(i,j;\tau)$ defined by
\begin{eqnarray}
{\bf G}_{2}(i\alpha,i\gamma,j\beta;\tau)
=\left( \begin{array}{cc}
u\langle \tau c^{\alpha}_{i \uparrow}(\tau){c^{\gamma}_{i\downarrow }(\tau)}^{\dagger}c^{\gamma}_{i\downarrow }(\tau){c^{\beta}_{j \uparrow}}^{\dagger}(0)\rangle
&
u\langle \tau c^{\alpha}_{i \uparrow}(\tau){c^{\gamma}_{i\downarrow }}^{\dagger}(\tau)c^{\gamma}_{i\downarrow }(\tau){c^{\beta}_{j \downarrow}}^{\dagger}(0)\rangle

\\
u\langle\tau c^{\alpha}_{i \downarrow}(\tau){c^{\gamma}_{i\uparrow }}^{\dagger}(\tau)c^{\gamma}_{i\uparrow }(\tau){c^{\beta}_{j \uparrow}}^{\dagger}(0)\rangle
&
u\langle\tau c^{\alpha}_{i \downarrow}(\tau){c^{\gamma}_{i\uparrow }}^{\dagger}(\tau)c^{\gamma}_{i\uparrow }(\tau){c^{\beta}_{j \downarrow}}^{\dagger}(0)\rangle
\end{array}\right).
\label{eq:two-particle-normal-general-1band}
\end{eqnarray}
The single particle equation of motion in an effective medium of Eq.\ref{eq:normal-general equation of motion} is 
\begin{eqnarray}
\sum_{l\gamma }\left((-{\bf I}\frac{\partial}{\partial\tau}+\mu{\bf I})\delta_{il}\delta_{\gamma\alpha}+{\bf t}^{\alpha\gamma}_{il}\right){\bar{\bf G}}^{\gamma\beta}(l,j;\tau)-\sum_{l\gamma}{\boldsymbol \Sigma}^{\alpha\gamma}(i,l;\tau){\bar{\bf G}}^{\gamma\beta}(l,j;\tau)
=\delta(\tau)\delta_{ij}\delta_{\alpha\beta}{\bf I}
\label{eq:normal-general equation of motion}
\end{eqnarray}
where self energy matrix ${\boldsymbol \Sigma}^{\alpha\gamma}(i,l;\tau)$ defined by
\begin{equation}
\sum_{\gamma}\langle {\bf G}_{2}(i\alpha,i\gamma,j\beta;\tau)\rangle=\sum_{l\gamma}{\boldsymbol \Sigma}^{\alpha\gamma}(i,l;\tau){\bar{\bf G}}^{\gamma\beta}(l,j;\tau)
\label{eq:self enrgy definition}
\end{equation}
From Eqs.\ref{eq:normal-general equation of motion} and \ref{eq:normal-general equation of motion} relation between spin-real space Green functions matrix are
\begin{equation}
{\bf G}(\tau)={\bar {\bf G}}(\tau) +{\bar {\bf G}}(\tau)\left ({\bf G}_{2}(\tau)- {\boldsymbol \Sigma}{\bf G}(\tau)\right )
\label{eq:G-G2}
\end{equation}
By taking average over all super cell except central super cell defined by $\{I, J\in sc\}$ Eq.\ref{eq:G-G2} reduces to\cite{Moradian03}
\begin{equation}
{\bf G}^{imp}_{sc}(\tau)={\bar {\bf G}}_{sc}(\tau) +{\bar {\bf G}}_{sc}(\tau)\left ({\bf G}^{imp}_{2 sc}(\tau)- {\boldsymbol \Sigma}_{sc}{\bf G}^{imp}_{sc}(\tau)\right )
\label{eq:G-G2-1}
\end{equation}
Eq.\ref{eq:G-G2-1} could be written as
\begin{equation}
{{\bar{\bf G}}_{sc}}^{-1}(\tau)+{\boldsymbol \Sigma}_{sc}=( {\bf G}^{imp}_{2 sc}(\tau) +{\bf I})  {{\bf G}^{imp}_{sc}(\tau)}^{-1}={\boldsymbol {\mathcal G}_{sc}}^{-1}    
\label{eq:G-G2-2}
\end{equation}
Eq.\ref{eq:G-G2-2} could be separated to two following equations
\begin{equation}
{{\bar{\bf G}}_{sc}}(\tau)={\boldsymbol {\mathcal G}}_{sc}+{\boldsymbol {\mathcal G}}_{sc} {\boldsymbol \Sigma}_{sc}{{\bar{\bf G}}_{sc}}(\tau)
\label{eq:G-G2-3}
\end{equation}
and
\begin{equation}
{\bf G}^{imp}_{sc}(\tau)={\boldsymbol {\mathcal G}}_{sc}+{\boldsymbol {\mathcal G}}_{sc}{\bf G}^{imp}_{2sc}(\tau)
\label{eq:G-G2-4}
\end{equation}
By quantum Monte Carlo method it is possible to convert two particle Green function in terms of Ising like fields\cite{Jarrell2001-2}
\begin{eqnarray}
 {\bf G}^{imp}_{sc}(i\omega_{n})= \boldsymbol{\mathcal{G}}(i\omega_{n})+\boldsymbol{\mathcal{G}}(i\omega_{n})\mathcal{V}_{\{s_{0},...,s_{4}\}} {\bf G}^{imp}_{sc}(i\omega_{n})
\label{eq:w7}
\end{eqnarray}
By imaginary time Fourier transform of Eq.\ref{eq:w7} we have
\begin{eqnarray}
 {\bf G}^{imp}_{sc}(\tau)= \boldsymbol{\mathcal{G}}(\tau)+\int d\tau^{'}\boldsymbol{\mathcal{G}}(\tau-\tau^{'})\mathcal{V}_{\{s_{0},...,s_{4}\}} {\bf G}^{imp}_{sc}(\tau^{'})
\label{eq:w8}
\end{eqnarray}
Average over random and Ising fields of super cell Green function $ {\bf G}_{sc}(\tau)$ gives us average Green function  ${\bar {\bf G}}(\tau)$
\begin{eqnarray}
\langle {\bf G}^{imp}_{sc}(\tau) \rangle= {\bar {\bf G}}(\tau)
\label{eq:w9}
\end{eqnarray}
Inverse Fourier transform of Eq.\ref{eq:w9} given by
\begin{eqnarray}
{\bar {\bf G}}(i\omega_{n})=\int d\tau \langle {\bf G}_{sc}(\tau) \rangle e^{i\omega_{n}\tau}
\label{eq:w10}
\end{eqnarray}
By analytical continuation of Eq.\ref{eq:w10} average Green function in real energy ${\bar {\bf G}}(E+i\eta)$ obtains. Process of extracting band structure from calculated ${\bar {\bf G}}(E+i\eta)$ and $\Sigma({\bf K}_{n}; E+i\eta)$ are as follows. The exact single particle effective Green function is 
\begin{eqnarray}
{\bar G}({n\bf k}; E+i\eta) &=& \frac{1}{(G^{0}({n\bf k}; E+i\eta))^{-1}-\Sigma({\bf k}; E+i\eta)}\nonumber\\&=&\frac{E-E^{0}_{n\bf k}-Re\Sigma({\bf k}; E+i\eta)-i(\eta-Im\Sigma({\bf k}; E+i\eta))}{(E-E^{0}_{n\bf k}-Re\Sigma({\bf k}; E+i\eta))^{2}+(\eta-Im\Sigma({\bf k}; E+i\eta))^{2}}
\label{eq:k-g-10}
\end{eqnarray}
where $E^{0}_{n \bf k}$ is non interacting bands. On the other hand  relation between exact effective Green function ${\bar G}({n\bf k}; E+i\eta)$ and effective band structure $E_{n\bf k}$ is 
\begin{eqnarray}
{\bar G}({n\bf k}; E+i\eta) &=& \frac{1}{E-E_{n\bf k}+i\eta}=\frac{E-E_{n\bf k}-i\eta}{(E-E_{n\bf k})^{2}+\eta^{2}}.
\label{eq:k-g-11}
\end{eqnarray}
For an exact effective medium system with whole lattice sites $lim_{\eta\rightarrow 0}Im\Sigma({\bf k}; E+i\eta)\rightarrow 0$, so effective band structures $E_{n\bf k}$ obtain from poles of effective Green function Eq.\ref{eq:k-g-10} 
\begin{eqnarray}
E=E^{0}_{n\bf k}+Re\Sigma({\bf k}; E+i\eta).
\label{eq:k-g-12}
\end{eqnarray}
In general, effective medium self energy $\Sigma({\bf k}; E+i\eta)$ can not be calculate exactly. Single site dynamical mean field approximation (dmft) with k-independent self energy  $\Sigma({\bf k}; E+i\eta)=\Sigma( E+i\eta)$ is lower approximation. Self energy in the cluster sites approximations such as dynamical cluster approximation (DCA) and effective medium super cell approximation (emsca) are step functions, $\Sigma({\bf K}_{m}; E+i\eta)$, that inside each $m$ grain in the first Brillouin zone is continuous but at grain boundaries are discontinuous. However these self energies discontinuities makes it impossible to calculate renormalized band structure.  Although density of states could be calculate from calculated local Green function $N(E)=-\frac{1}{\pi}\ Im {\bar G}(I,I; E+i\eta)$. Another important problem of these approximations is creating fake electronic states which leads to $lim_{\eta\rightarrow 0}Im\Sigma({\bf k}; E+i\eta)\neq 0$. One expect by increasing number of sites in the cluster number of fake states should decrease. By help of real states density of states 
\begin{eqnarray}
N_{real}({\bf k}; E) =\frac{1}{\pi}\sum_{n} \frac{\eta}{(E-E_{n\bf k})^{2}+\eta^{2}}.
\label{eq:k-g-14}
\end{eqnarray}
that at $E=E_{n\bf k}$ has a maximum like a Dirac delta function one can extract real state eigen values and eliminate fake states. So we should find maximums of calculated density of states, $N({\bf k}; E)$, 
\begin{eqnarray}
N({\bf k}; E) &=& \frac{1}{\pi} \sum_{n} \frac{\eta-Im\Sigma({\bf k}; E+i\eta)}{(E-E^{0}_{n\bf k}-Re\Sigma({\bf k}; E+i\eta))^{2}+(\eta-Im\Sigma({\bf k}; E+i\eta))^{2}}.
\label{eq:k-g-13}
\end{eqnarray}
that could be obtain from
\begin{eqnarray}
\frac{d N({\bf k}; E)}{dE} =0,\;\; \frac{d^{2} N({\bf k}; E)}{d^{2}E}> 0.
\label{eq:k-g-013}
\end{eqnarray}
 Fig.\ref{figure:real-dos}  shows this.  To reveal advantage of our method we calculate renormalized graphene band structure in dmft and four sites, $nc=4$,  beyond super cell approximation. 
 First we applied dynamical mean field theory to the two bands graphene like lattice for different on site electrons repulsions at half band filling $n=n_{\uparrow}+n_{\downarrow}=1$, $n_{\uparrow}=0.5$ and $n_{\downarrow}=0.5$. Our results show that by increasing electrons energy repulsion $u$ valance and conduction bands separated but steel valance band completely full by both spin up and down electrons and conduction band is empty.. Fig.\ref{figure:dmft-u2-56}  shows calculated realistic two bands for $u=2.56 t$ in which the fake states are eliminated. To high light advantage of our method we compared the direct dmft calculated density of states $N({\bf k}; E)$ with density of states obtained from calculated realistic valance and conduction bands. We see that there are some fake states in the direct obtained dmft dos.  

\begin{figure}
\centerline{\epsfig{file=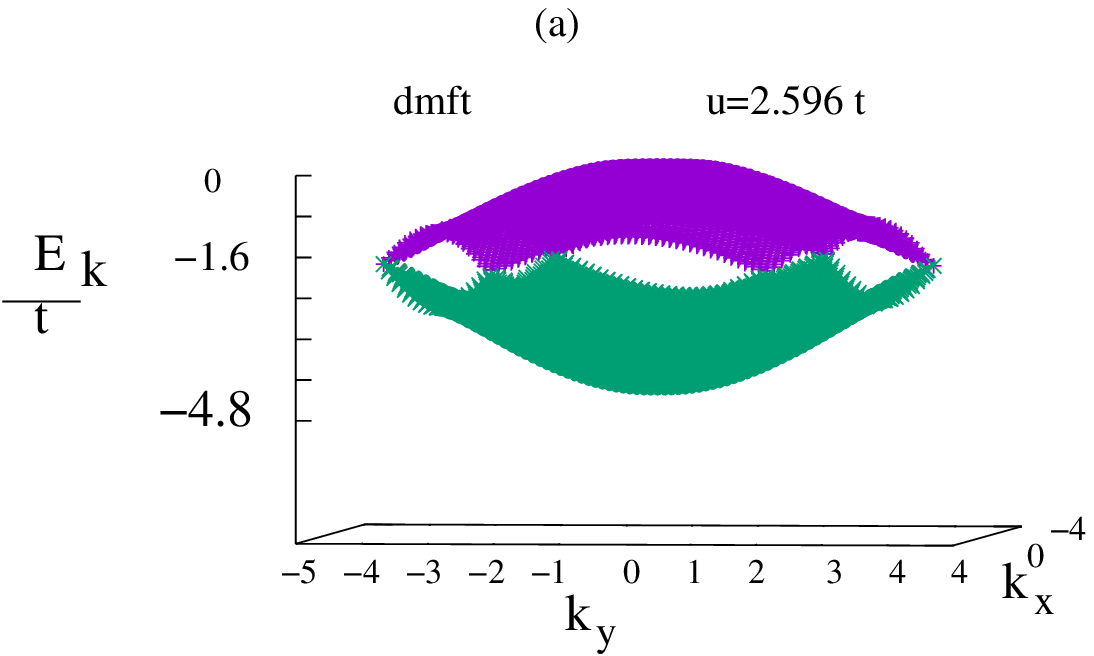  ,width=6.0cm,angle=0}\epsfig{file= 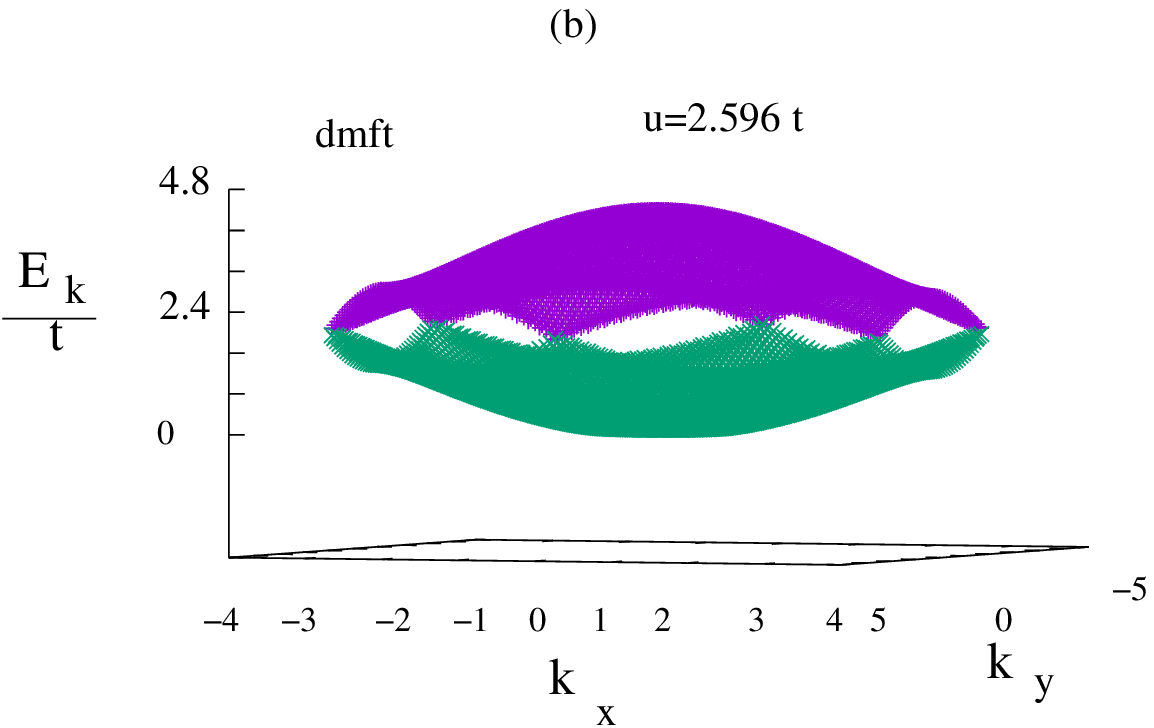 ,width=6.0cm,angle=0}}
\centerline{\epsfig{file=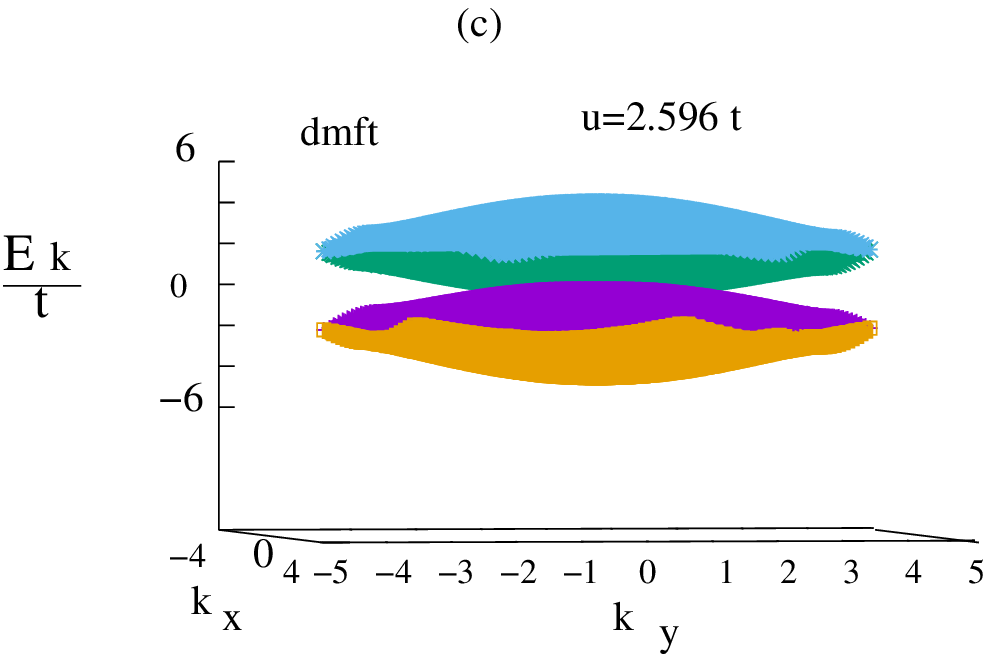 ,width=6.0cm,angle=0}\epsfig{file=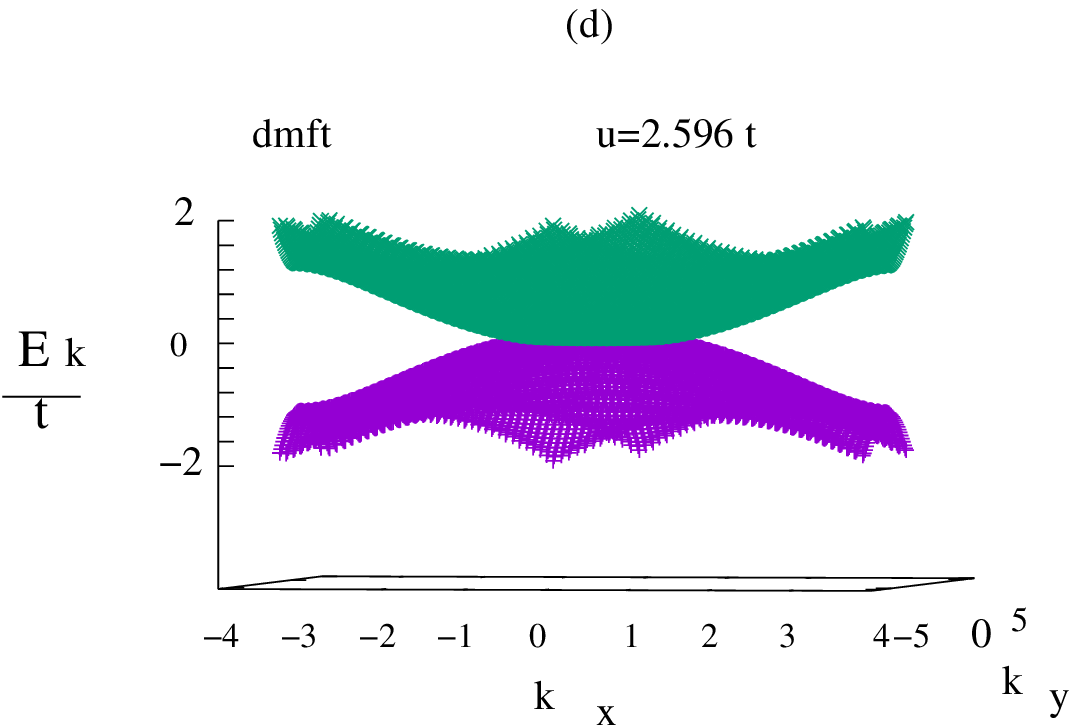 ,width=6.0cm,angle=0}}
\centerline{\epsfig{file= 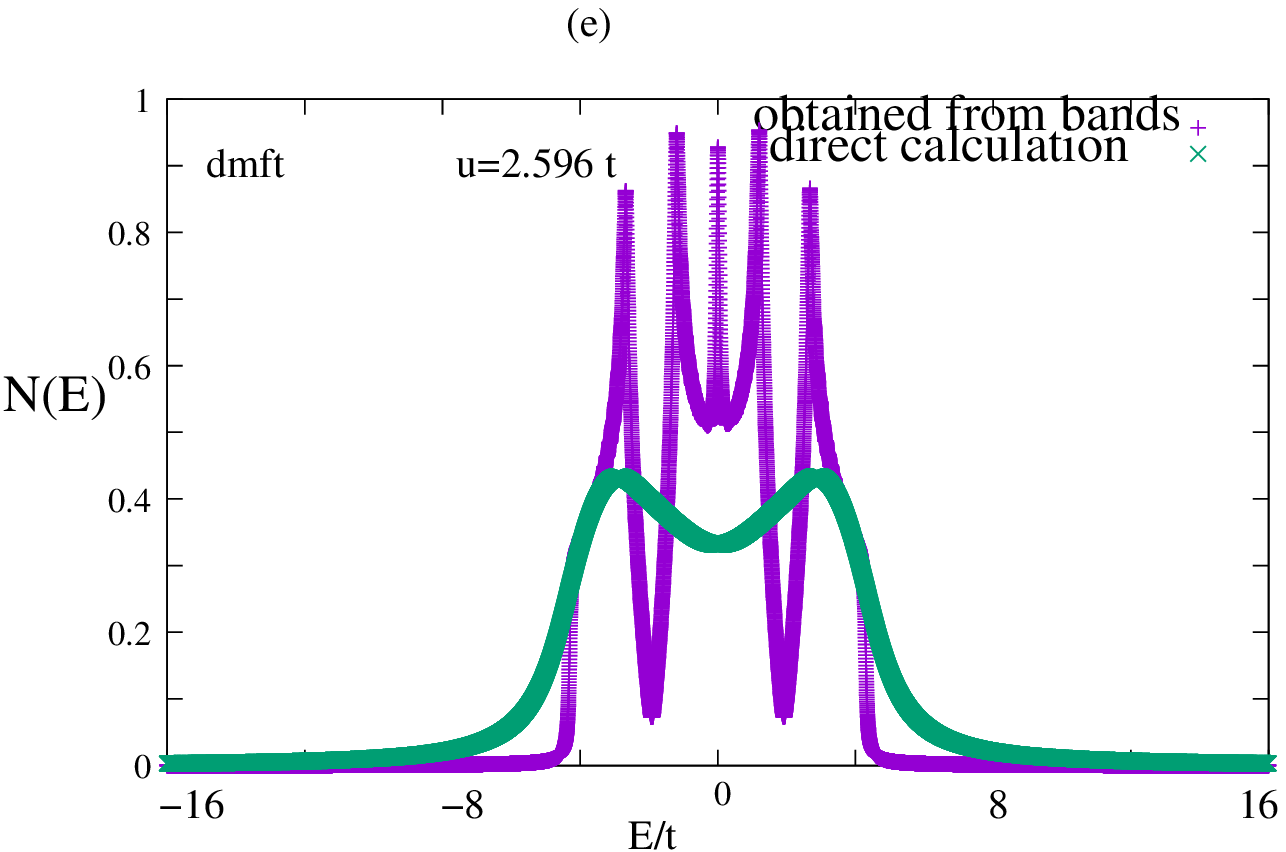  ,width=6.0cm,angle=0}}
\caption{(Color on line) (a), (b), (c) and (d) shows realistic dmft calculated bands for critical repulsion energy $u=2.596 t$. (e) is whole bands together. (f) shows comparison of dos obtained from realistic bands with dos obtained from calculated local Green function.  }
 \label{figure:dmft-u2-596} 
\end{figure}

In the dmft the critical value of repulsion for separating four bands with zero gap at Fermi energy is $u=2.596 t$. Two of bands touch Fermi energy at different ${\bf k}$ points. For $u>2.596 t$ we have four separated bands with an energy gap at Fermi energy. Two of these bands are spin up and other two bands belong to spin down electrons. This two bands to four bands spin up and down with gap at Fermi energy called semimetal to anti ferromagnetic Mott transition. Fig.\ref{figure:dmft-u2-596}  (a), (b), (c) and (d) show dmft calculated bands for $u=2.596 t$ and (e) illustrate whole bands. Fig.\ref{figure:dmft-u2-596} (f) shows comparison of dos obtained from realistic calculated bands by eliminating fake states and dos obtained from dmft local single particle Green function.  

\begin{figure}
\centerline{\epsfig{file= 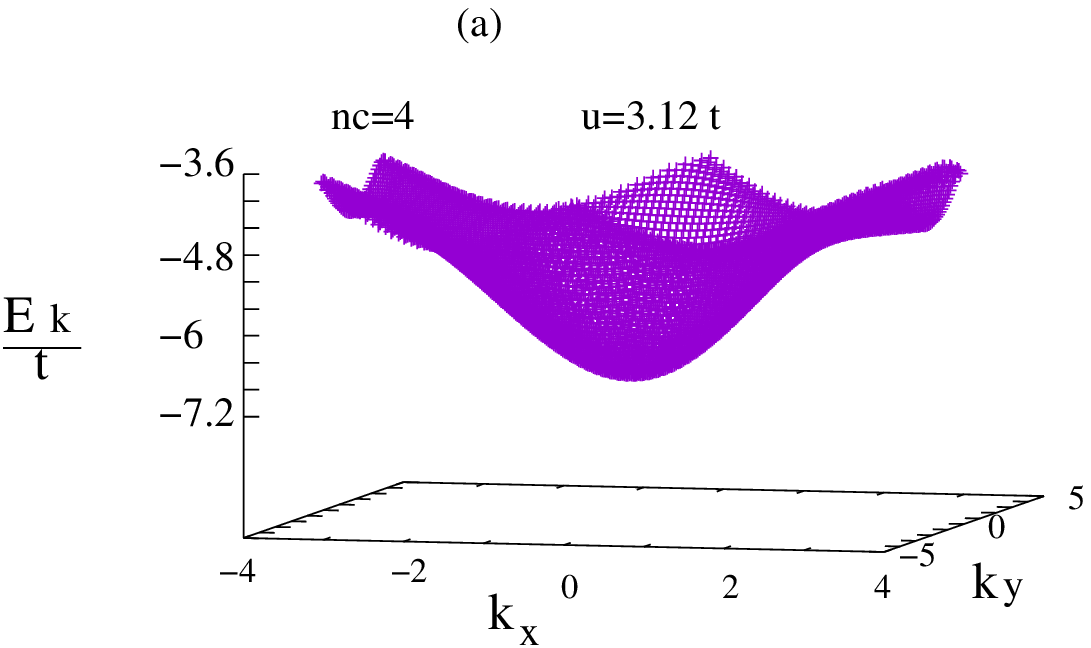,width=6.0cm,angle=0}\epsfig{file= 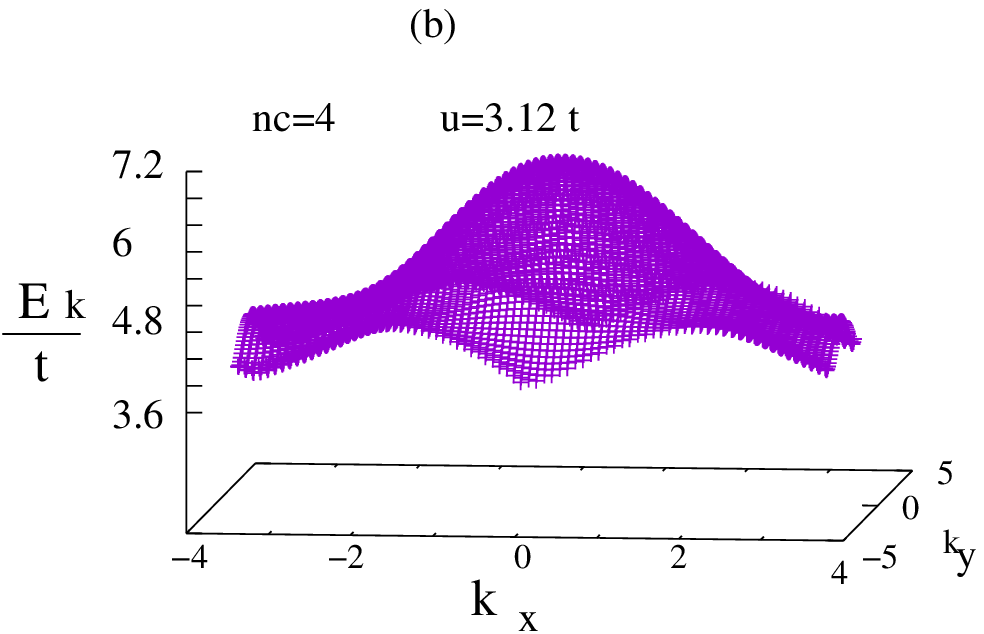 ,width=6.0cm,angle=0}}
\centerline{\epsfig{file= 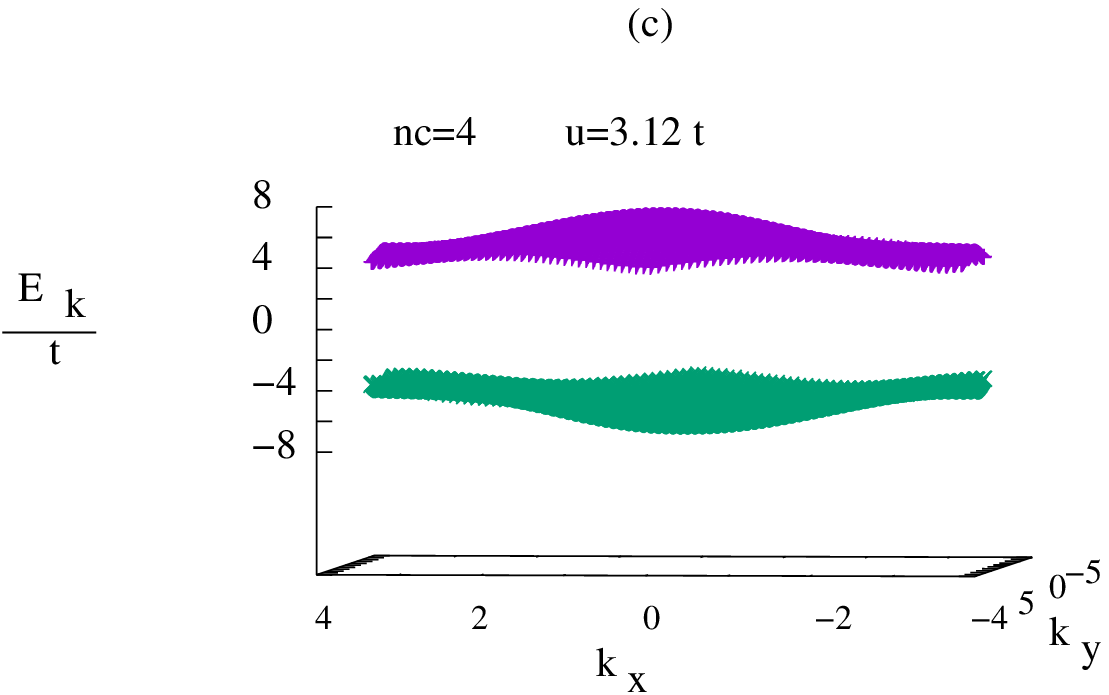
,width=6.0cm,angle=0}\epsfig{file=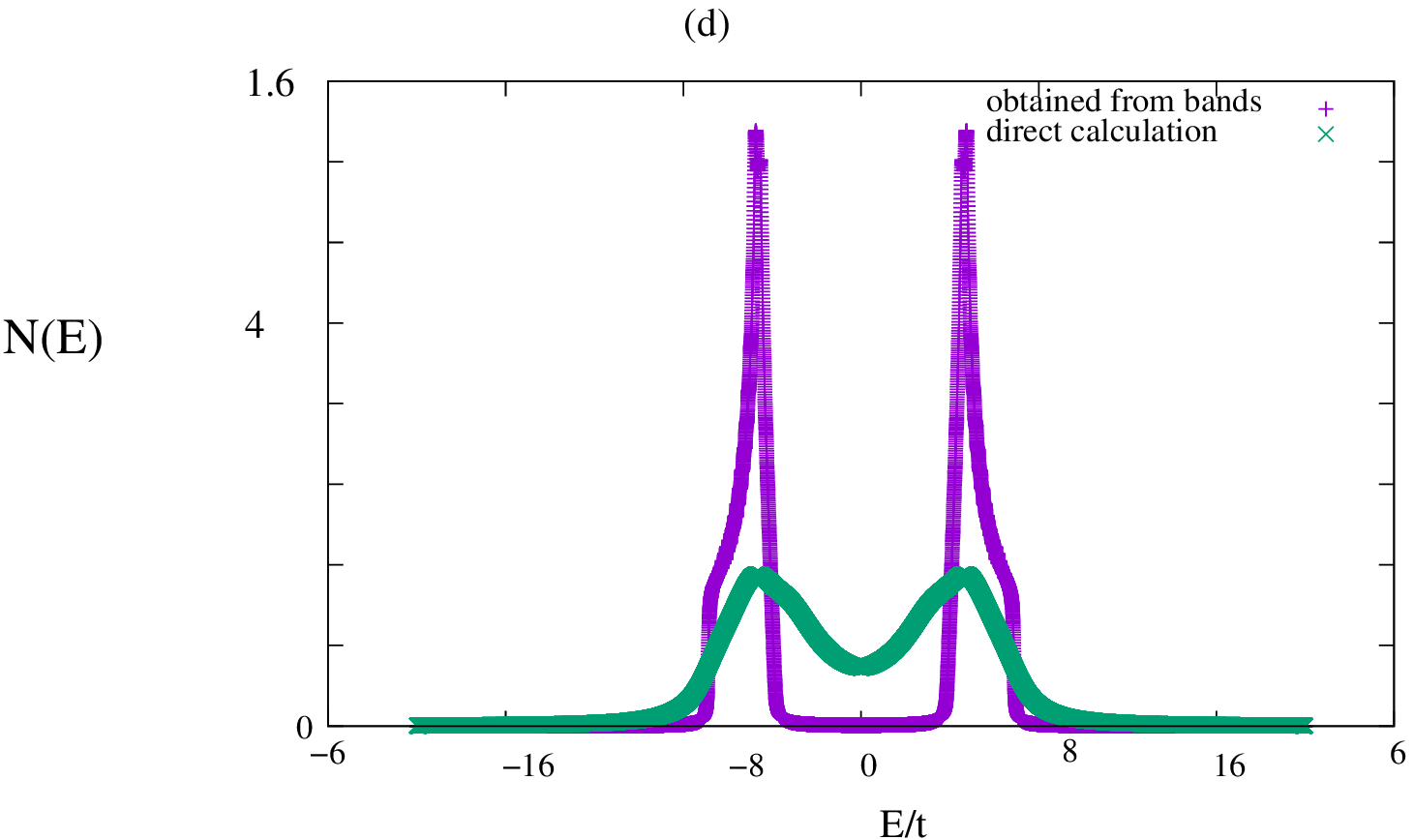 ,width=6.0cm,angle=0}}
\caption{(Color on line) (a), (b), (c) and (d) shows four realistic spin up and down calculated bands in the beyond  $nc=4$ super cell approximation for $u=3.12 t$. (e) illustrate whole bands. (f) shows dos calculated from realistic bands and compared with dos from local Green function in $nc=4$ approximation.  }
 \label{figure:nc4-u3-12} 
\end{figure}

To see effects of multi sites correlation on band structure and density of states first we applied four sites $nc=4$ effective medium super cell approximation to obtain super cell self energy $\Sigma(I, J;E)$ and $\Sigma({\bf K}_{n}; E)$\cite{Moradian01,Moradian02,Moradian03}. Then we approximate ${\bf k}$-space self energy that is continuous in the first Brillouin zone by $\Sigma({\bf k}; E)=\frac{1}{Nc}\sum_{IJ} e^{i {\bf k}.{\bf r}_{IJ}} \Sigma(I, J; E)$. By substitution this calculated  $\Sigma({\bf k}; E)$ in Eq.\ref{eq:k-g-13} and using Eq.\ref{eq:k-g-14} all bands calculated. Fig.\ref{figure:nc4-u3-12} (a), (b) show realistic bands for $u=3.12 t$. (c) is whole bands and (d) shows comparison of dos obtained from realistic calculated bands and dos obtained directly from calculated $nc=4$ super cell Green function. Our results show that in this repulsion energy spin up and down bands are not separated.

\begin{figure}
\centerline{\epsfig{file=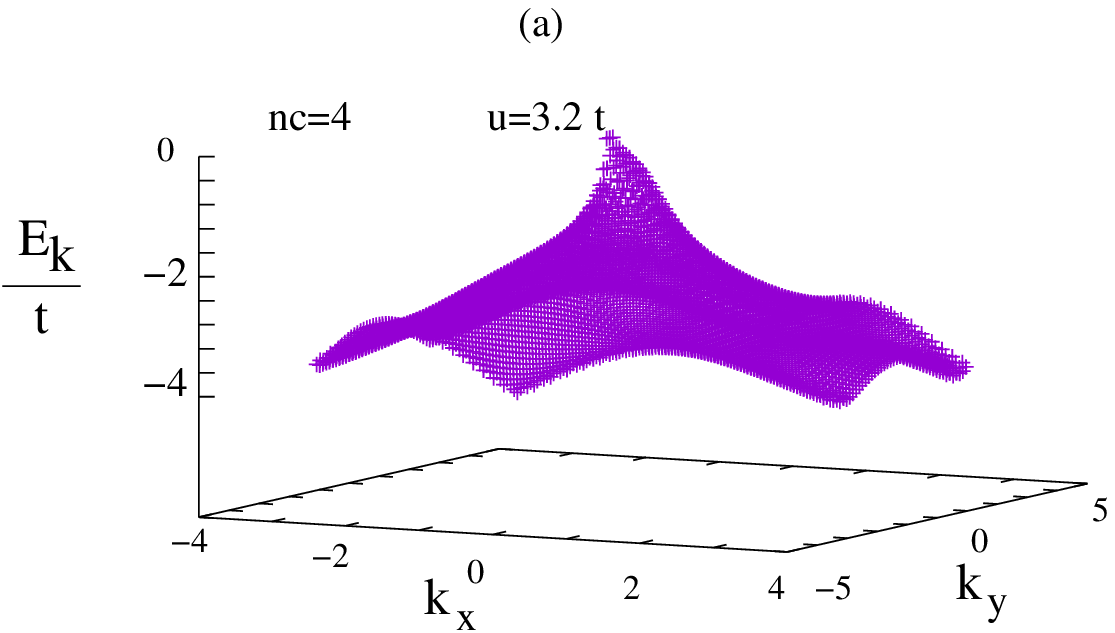  ,width=6.0cm,angle=0}\epsfig{file= 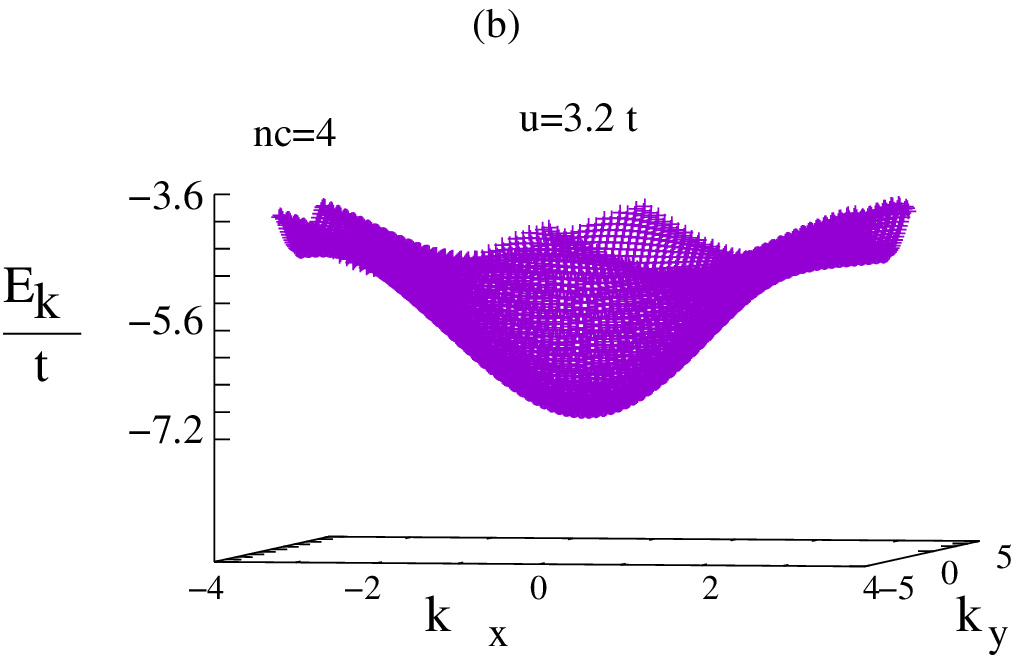 ,width=6.0cm,angle=0}}
\centerline{\epsfig{file=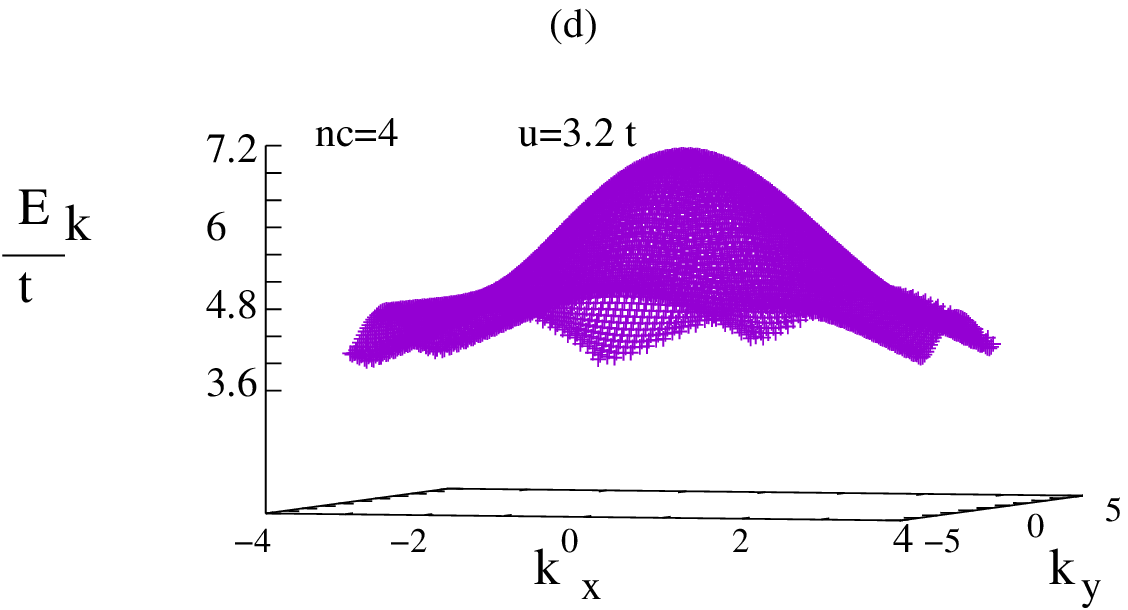 ,width=6.0cm,angle=0}\epsfig{file= 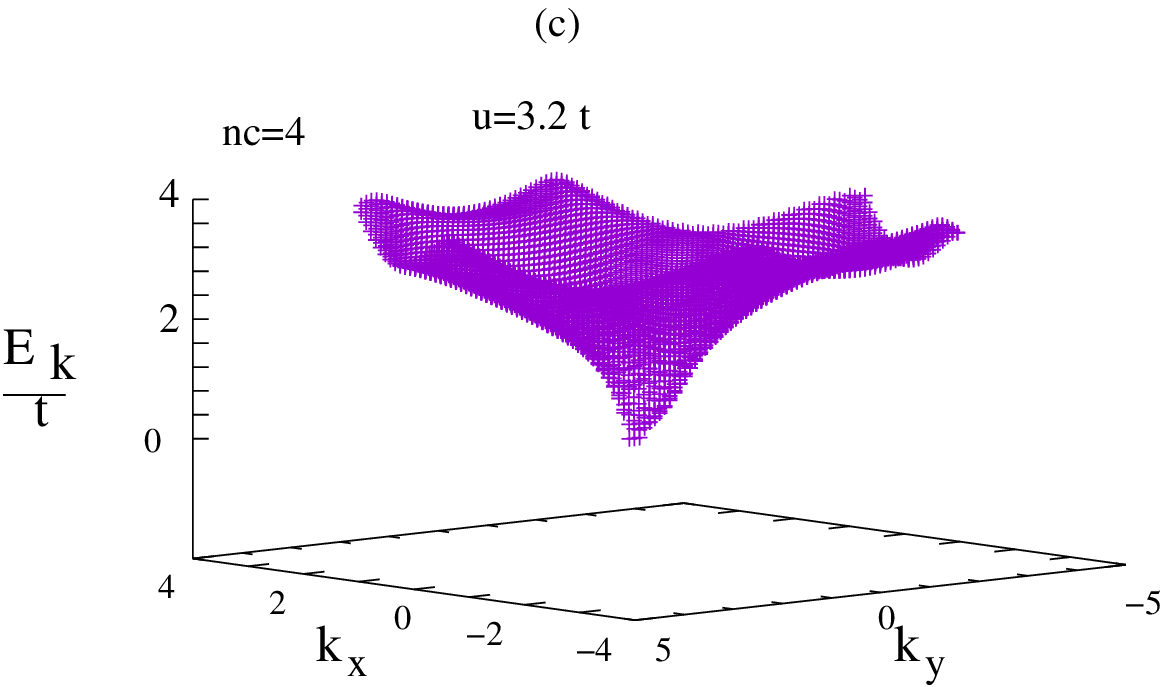,width=6.0cm,angle=0}}
\centerline{\epsfig{file=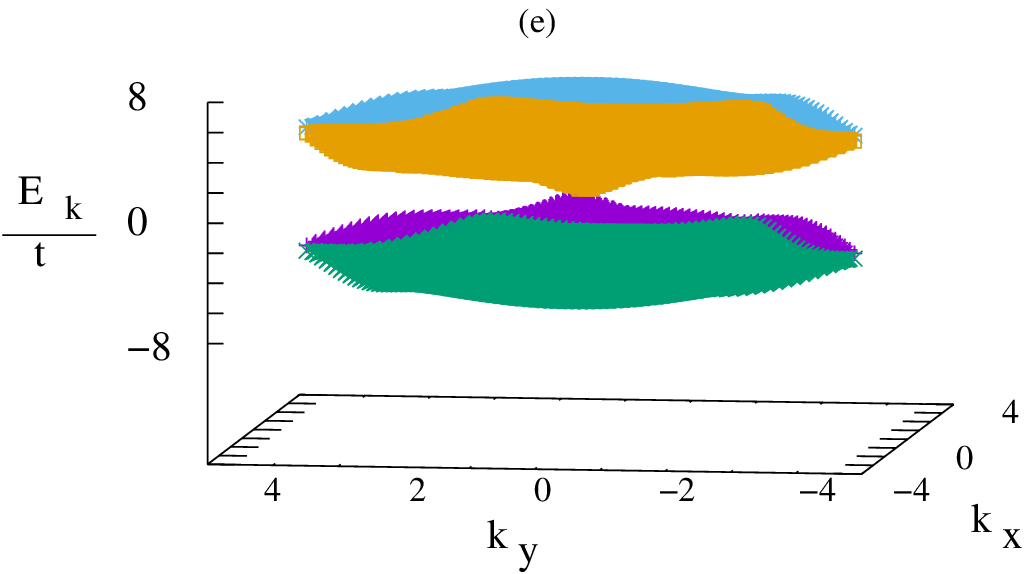   ,width=6.0cm,angle=0}\epsfig{file= 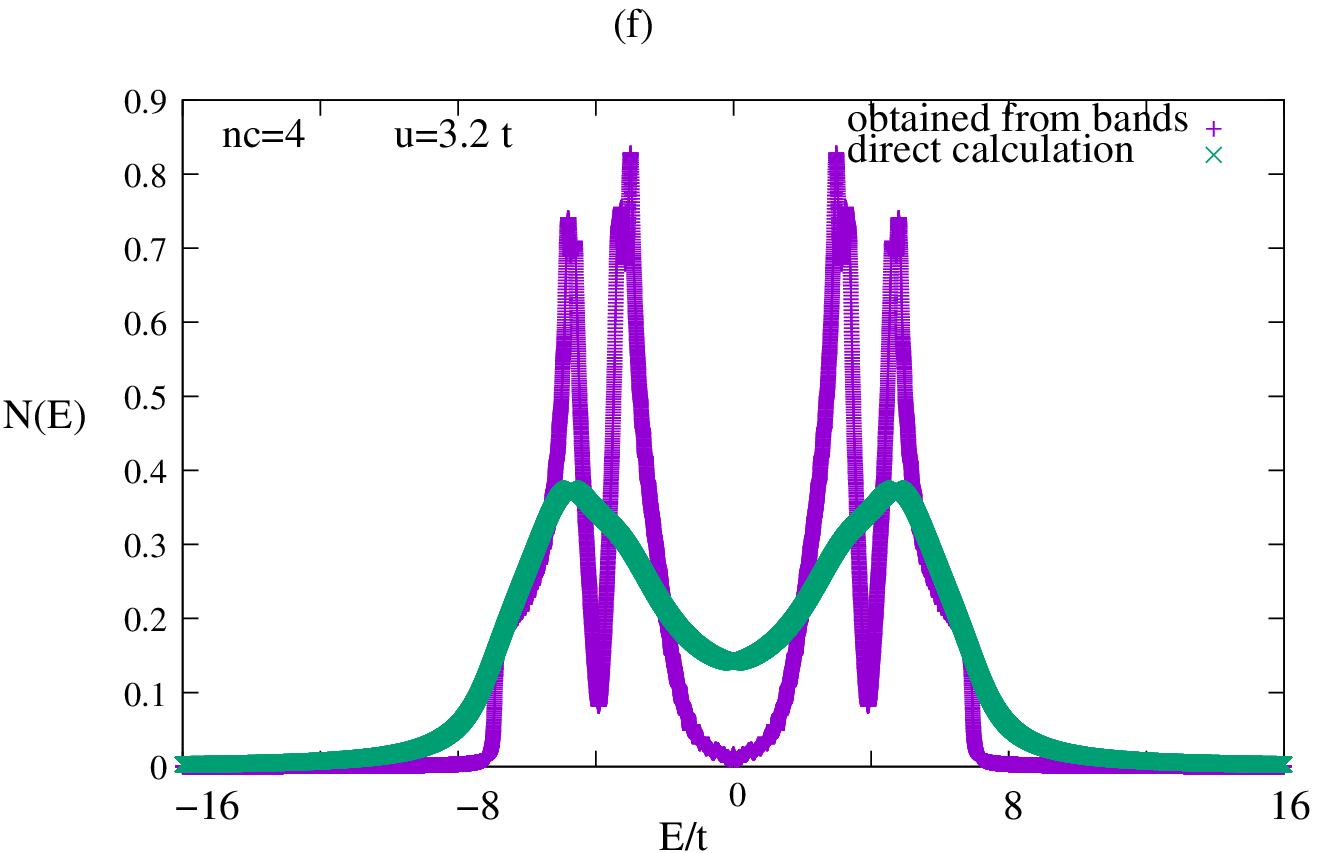  ,width=6.0cm,angle=0}}
\caption{(Color on line) (a), (b), (c) ,(d) shows realistic calculated spin up and down electron bands for $u=3.12 t$. (e) illustrate whole realistic bands and (f)  shows comparison of realistic dos and dos obtained directly from local Green function obtained from $nc=4$ approximation.}
 \label{figure:nc4-u3-2} 
\end{figure}

Our results show that at critical repulsion potential $u=3.2 t$ spin up and down electron separate in which valance and conduction bands touch Fermi energy at two different ${\bf k}$ points. Fig.\ref{figure:nc4-u3-2}  (a), (b), (c) and (d) shows calculated realistic bands in the $nc=4$ approximation for $u=3.2 t$. (e) illustrate whole bands. (f) shows comparison of calculated dos from realistic bands and calculated directly from local Green function. 

In conclusion a realistic renormalized band structure of Hubbard model of an interacting electrons graphene like lattice obtained. Until now people taught that all states obtained by approximated self energies are acceptable but we showed that this is not correct. This completely changes physics of  results.

\end{document}